\documentclass{LT23auth}
\usepackage{graphicx}
\usepackage{amssymb}
\usepackage{amsfonts}

\begin{document}

\begin{frontmatter}

\title{Effects of disorder on conductance 
through small interacting systems}

\author{Yoshihide Tanaka\thanksref{thank1}}
\author{and Akira Oguri}

\address[address1]{Department of Material Science, Faculty of Science, 
Osaka City University, Sumiyoshi-ku, Osaka 558-8585, Japan}

\thanks[thank1]{E-mail: tanakay@sci.osaka-cu.ac.jp}

\begin{abstract}
We study the effects of disorders on the transport 
through small interacting systems based on 
a two-dimensional Hubbard cluster 
of finite size connected to two noninteracting leads. 
This system can be regarded as a model 
for the superlattice of quantum dots or 
atomic network of the nanometer size. 
We calculate the conductance at $T=0$ 
using the order $U^2$ self-energy in an electron-hole symmetric case. 
The results show that the conductance is sensitive to the randomness 
when the resonance states are situated near the Fermi energy. 
\end{abstract}


\end{frontmatter}

Quantum transport through small interacting systems, 
such as quantum dots and wires,
has been a subject of current interest. 
For these systems theoretical approaches, which are able to treat 
correctly the interaction and interference effects, are necessary 
for systematic investigations. 
In this report, using the perturbation method described in \cite{YT}, 
we study effects of disorders 
on the conductance of the interacting systems. 
Effects of disorders seem to be important
in the systems consisting of a number of
quantum dots,  because randomness must be
inevitable for artificially fabricated systems.

We consider a system consisting of three regions; 
a finite interacting region at the center ($\mathrm{C}$), 
and two noninteracting reservoirs on the left ($\mathrm{L}$) and 
right ($\mathrm{R}$). 
The total Hamiltonian is, 
$
\mathcal{H}_{\mathrm{tot}}  
=   \mathcal{H}^0 + \mathcal{H}_{\mathrm{C}}^{\mathrm{int}},
$
with 
$
\mathcal{H}^0  
=  \mathcal{H}_{\mathrm{L}} + \mathcal{H}_{\mathrm{R}}  
+ \mathcal{H}_{\mathrm{C}}^0 + 
\mathcal{H}_{\mathrm{mix}} 
$. 
Here 
$\mathcal{H}_{\mathrm{L}}$ ($\mathcal{H}_{\mathrm{R}}$) is 
a Hamiltonian for the left (right) lead. 
The central region is described by 
a Hubbard model on a square lattice 
of $N\times M$ ($=N_{\mathrm{C}}$) sites: 
$
\mathcal{H}_{\mathrm{C}}^0 = -
\sum_{jj' \in \mathrm{C}}\, t_{jj'}^{\mathrm{C}}
 c^{\dagger}_{j \sigma} c^{\phantom{\dagger}}_{j' \sigma}  
$, and  
$
\mathcal{H}_{\mathrm{C}}^{\mathrm{int}} = U \sum_{j=1}^{N_{\mathrm{C}}} 
\left[\, 
n_{j \uparrow}\, n_{j \downarrow}-(n_{j \uparrow}
+ n_{j \downarrow} )/2 \, \right] 
$ in the standard notation. 
In this report we consider 
the electron-hole symmetric case, and 
examine effects of the off-diagonal disorder 
described the nearest-neighbor transfer $t_{jj'}^{\mathrm{C}}$. 
The central region and two leads are connected 
via $M$ channels described by $\mathcal{H}_{\mathrm{mix}}$ \cite{YT}. 
We assume that the two coupling are equal and described by a parameter 
$\Gamma = \pi \rho\, v^2 $, 
where $v$ is the mixing matrix element 
and $\rho$ is the density of states of the isolated leads. 
The system may be regarded as a model 
for the superlattice of quantum dots \cite{tamura}. 

The zero-temperature conductance of this system 
is determined by the value of the single-particle Green's function 
at the Fermi energy $\omega=0$ \cite{YT}. 
The Dyson equation is written 
in a $N_{\mathrm{C}}\times N_{\mathrm{C}}$ matrix form 
\begin{eqnarray}
\left\{ \widehat{\mbox{\boldmath $\mathcal{G}$}}(z) \right\}^{-1} &=& 
\, 
\left\{ \widehat{\mbox{\boldmath $\mathcal{G}$}}^0(z) \right\}^{-1} 
 - \widehat{\mbox{\boldmath $\Sigma$}}(z),
\label{eq:G}
\\
\left\{ \widehat{\mbox{\boldmath $\mathcal{G}$}}^0(z) \right\}^{-1} 
&=&\, 
z \, \widehat{\mbox{\boldmath $1$}}
 - \widehat{\mbox{\boldmath $\mathcal{H}$}}_{\mathrm{C}}^0 
- \widehat{\mbox{\boldmath $\mathcal{V}$}}_{\mathrm{mix}}(z) \;. 
\label{eq:G0}
\end{eqnarray}
Here 
$\widehat{\mbox{\boldmath $\mathcal{G}$}}^0
= \left\{ G_{jj'}^{0} \right\}$ 
with $j j' \in \mathrm{C}$ is 
the unperturbed Green's function corresponding to $\mathcal{H}^0$, and 
$\widehat{\mbox{\boldmath $\Sigma$}} 
= \left\{ \Sigma_{jj'} \right\}$ 
is the self-energy 
due to the interaction $\mathcal{H}_{\mathrm{C}}^{\mathrm{int}}$. 
We assume the hard-wall boundary condition for 
$\widehat{\mbox{\boldmath $\mathcal{H}$}}_{\mathrm{C}}^0 
 = \left\{ t_{jj'}^{\mathrm{C}} \right\}$ 
along the direction perpendicular to the current. 
The size  along this direction is $M$. 
The mixing self-energy 
$\widehat{\mbox{\boldmath $\mathcal{V}$}}_{\mathrm{mix}}$ 
is non-zero only for the two subspaces corresponding to the interfaces, 
for each of which the partitioned matrix is 
given by $-i \Gamma \mbox{\boldmath $1$}$ for 
the retarded function $z=\omega+i0^+$ 
with $\mbox{\boldmath $1$}$ being the unit matrix of size $M$. 
We calculate the value 
of $\widehat{\mbox{\boldmath $\mathcal{G}$}}$ at $\omega=0$ using 
eq.\ (\ref{eq:G}) 
with the 
order $U^2$ self-energy $\widehat{\mbox{\boldmath $\mathcal{\Sigma}$}}(0)$ 
as in \cite{YT}. 
In the present study, we take 
the nearest-neighbor transfer to be random variables between 
$0.9<t_{jj'}^{\mathrm{C}}/t<1.1 $, and 
take the strength of the mixing to be $\Gamma/t=0.75$, 
where $t$ is the transfer for the regular cluster. 

\begin{figure}[tb]
\begin{center}
\leavevmode
\includegraphics[width=1.0\linewidth, clip, 
trim = 0cm 14.7cm 0cm 1.2cm]{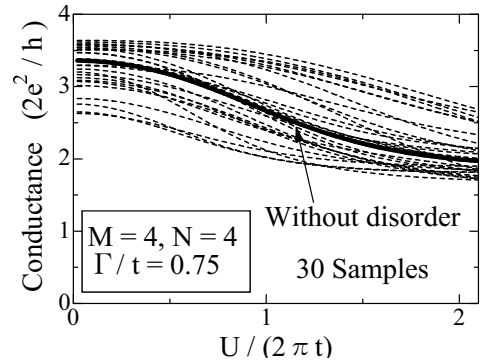}
\caption{Conductance of a Hubbard cluster of $4\times 4$ 
at half-filling with the off-diagonal randomness $0.9 <t'/t <1.1$. 
Disordered samples (29 dashed lines), and regular cluster (solid line). 
}
\label{trans1}
\end{center}
\end{figure}

\begin{figure}[tb]
\begin{center}
\leavevmode
\includegraphics[width=1.0\linewidth, clip, 
trim = 0cm 14.7cm 0cm 0cm]{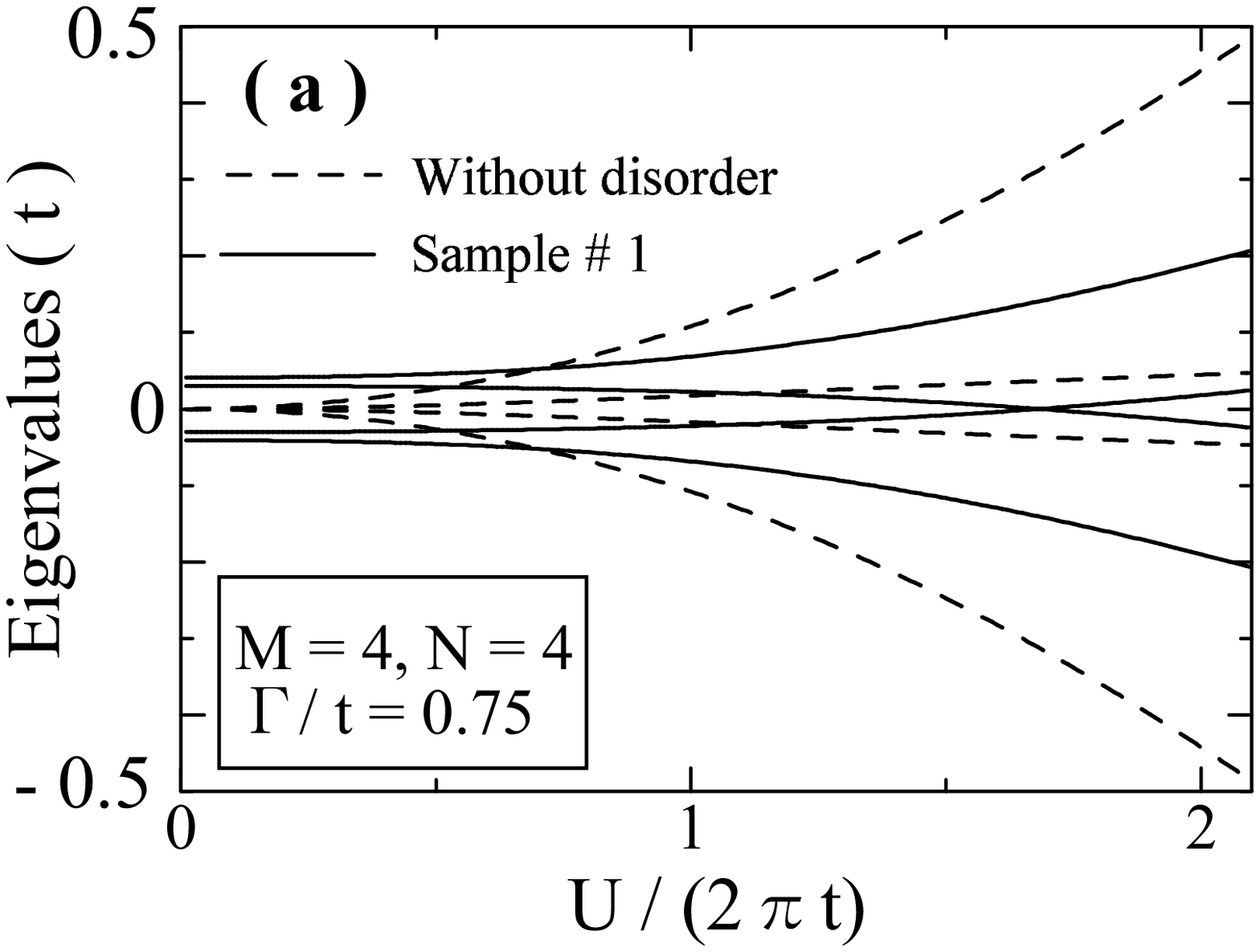}
\includegraphics[width=1.0\linewidth, clip, 
trim = 0cm 14.7cm 0cm 0cm]{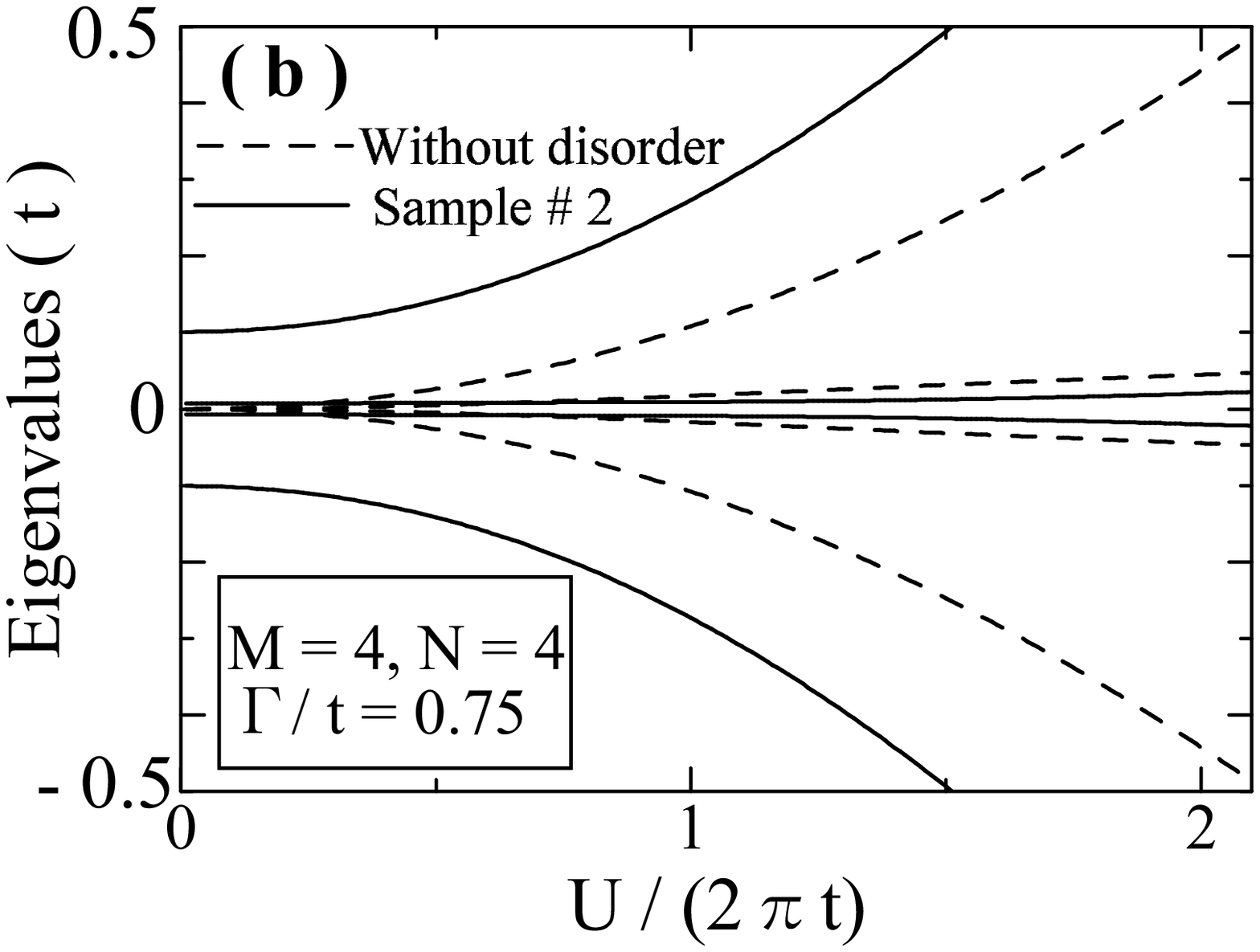}
\caption{Eigenvalues of the effective Hamiltonian 
$\widehat{\mbox{\boldmath $\mathcal{H}$}}_{\mathrm{C}}^{\rm eff}$ 
near the Fermi energy $\omega=0$, which are related to 
the resonance states contributing to the current. 
}
\label{eigen}
\end{center}
\end{figure}

We show the results obtained for the system 
of the size $M=4$ and $N=4$. 
In Fig.\ \ref{trans1} 
the conductance for 29 different samples of 
random configurations (dashed lines) and 
that of the regular cluster without the disorder (solid line) 
are plotted vs $U$. 
The conductance for each of these samples decreases with increasing $U$. 
Due to the randomness, the value of the conductance fluctuates 
around that of the regular cluster. 
To see these features of the results from a different viewpoint, 
we evaluate the eigenvalues of an effective Hamiltonian 
$
\widehat{\mbox{\boldmath $\mathcal{H}$}}_{\mathrm{C}}^{\rm eff} 
\equiv \widehat{\mbox{\boldmath $\mathcal{H}$}}_{\mathrm{C}}^0 
+ \widehat{\mbox{\boldmath $\Sigma$}}(0)$. 
The eigenvalues can be related to 
the peak position of the resonance states \cite{YT}, 
and among $N_{\mathrm{C}}$ eigenstates 
those near the Fermi energy $\omega=0$ contribute to the transport. 
In Fig.\ \ref{eigen} 
the eigenstates near $\omega=0$ are 
compared to those for the regular cluster without disorder (dashed lines), 
where the sample \#1 (\#2) is a typical example, 
the conductance of which is larger (smaller) 
than that of the regular cluster. 
Note that the eigenvalues are symmetric with respect 
to $\omega=0$ due to the electron-hole symmetry, 
and thus the four eigenvalues shown in each figure 
are classified into two pairs. 
In both of the samples 
the pair situated closer to the Fermi energy 
keep staying near $\omega=0$ for $U/(2\pi t) \lesssim 2.0$, 
while the other pair leave away from the Fermi energy 
as $U$ increases. 
Therefore, 
for the results of the conductance shown in Fig.\ \ref{trans1} 
the contribute of the pair at the near side 
is about $2\, (2e^2/h)$ and almost independent of the value of 
the onsite repulsion for small $U/(2 \pi t)$. 
Thus, the other pair at the far side mainly determine 
the $U$ dependence of the conductance of the cluster examined here. 

This feature can be understood from the electronic structure of 
the regular cluster for $U=0$, where the system has a well-defined 
subband structure. 
In this limit the pair at the far side correspond 
to the resonant states in the lowest and highest subbands, 
which are situated near the edge of the subbands. 
Since in each of these two subbands the Fermi energy is also close to 
the band edge, 
the position of the resonance states relative to the Fermi energy 
becomes sensitive to the disorder and interaction. 
The contributions of the conducting subbands at the marginal positions 
play an important role for the fluctuation of the current through 
the systems with a small number of conducting channels. 
Details of the formulation, numerical results and discussions will be 
presented elsewhere.

\medskip

We would like to thank H. Ishii for valuable discussions. 
Numerical computation was partly performed at computation 
center of Nagoya University and at Yukawa Institute Computer Facility. 
This work is supported by the Grant-in-Aid 
for Scientific Research from the Ministry of Education, 
Science and Culture, Japan.

\begin{thebibliography}{9}
\bibitem{YT}
Y. Tanaka, A. Oguri, and H. Ishii, 
J.\ Phys.\ Soc.\ Jpn.\ {\bf 71} (2002) 211.
\bibitem{tamura} 
H. Tamura, K. Shiraishi, and H. Takayanagi, 
J.\ Appl.\ Phys.,\ part2 {\bf 39} (2000) L241.
\end{thebibliography}
\end{document}